\documentclass[twocolumn,prb,showpacs]{revtex4}
\usepackage{graphicx}
\usepackage{dcolumn}
\usepackage{bm}

\begin{document}

\title{Influence of the magnetopolaron effect on light reflection and absorption by a wide semiconductor quantum well}
\author{I. G. Lang, L. I. Korovin}
\affiliation{A. F. Ioffe Physical-Technical Institute, Russian
Academy of Sciences, 194021 St. Petersburg, Russia}
\author{S. T. Pavlov\dag\ddag}
\address{\dag Facultad de Fisica de la UAZ, Apartado Postal C-580, 98060 Zacatecas, Zac., Mexico\\
\ddag P. N. Lebedev Physical Institute, Russian Academy of
Sciences, 119991 Moscow, Russia}

\begin{abstract}
Light reflection and absorption spectra by a semiconductor quantum
well (QW) , which width is comparable to a light wave length of
stimulating radiation, are calculated. A resonance with two close
located exited levels is considered. These levels can arise due to
splitting of an energy level of an electron-hole pair (EHP) due to
magnetopolaron effect, if the QW is in a quantizing magnetic field
directed perpendicularly to the QW plane. It is shown that unlike
a case of narrow QWs light reflection and absorption depend on a
QW width $d$. The theory is applicable at any ratio of radiative
and non-radiative broadenings of electronic excitations.
\end{abstract}


\maketitle

\section{Introduction}

At light transmission through a QW some characteristics occur in
reflected and transmitted light, on which it is possible to judge
about electronic processes proceeding in a QW [1-4]. Most
interesting results turn out when energy levels of electronic
system are discrete. It takes place in a quantizing magnetic field
directed perpendicularly to a QW plane or at taking into account
excitonic levels in a zero magnetic field.

Two close located energy levels  arise in a case
 of a magnetophonon  resonance [5], when
\begin{equation}
\label{1} \omega_{LO}=j\,\Omega,
\end{equation}
where $ \omega _{ LO} $ is the longitudinal optical $ (LO) $
phonon frequency, $j$ is the integer and
\begin{equation}
\label{2} \Omega = | e|H / cm _{ e (h)}
\end{equation}
is the cyclotron frequency,  $e$ is the electron charge, $m_{e
(h)}$ is the electron (hole) effective mass.

The modern semiconductor technologies allow to make QWs of high
quality, when radiative broadening of an absorption line may be
comparable to contributions of non-radiative mechanisms or to
exceed them. In such situation it is impossible to be limited by
linear interaction of an electron with electromagnetic field and
it is necessary to take into account all the orders of this
interaction [6-22].

Light pulses and monochromatic
 radiation were considered as stimulating light. One, two and large number of exited levels were taken into
account. Results of all previous works, except [19,20], are fair
for rather narrow QWs, when the condition
\begin{equation}
\label{3} \kappa d < < 1,
\end{equation}
is satisfied ($ \kappa $ is the module of a light wave vector, $d
$ is the QW width). As calculations show, in case of narrow QWs
only positions of reflection and absorption peaks depend on the
QW width, but not their height and form.

In [19,20] the theory of light reflection and absorption is
constructed for wider QWs for which
\begin{equation}
\label{4} \kappa d\ge 1.
\end{equation}
In both works the interaction of light with one exited level was
considered, in [19] - at a monochromatic irradiation, in [20] - at
a pulse irradiation. Under condition (4) results begin to depend
on the QW width $d $.

In the present work we investigate theoretically reflection,
absorption and transmission of monochromatic light through a wide
QW in a case, when the interaction of light with two close
located energy levels is essential. We compare new results to
conclusions of [21] devoted to research of similar problems in
 narrow QWs.

\section{Statement of a task and initial equations}

The case of normal incidence of light on a semiconductor QW
surface, located in a plane $xy $, is considered. QW may be in a
zero or quantizing magnetic field perpendicular to a QW surface.
Temperature is close to 0. Light excites electron-hole pairs
(EHPs). In the theory interband matrix elements ${\bf p}_{cv} $,
describing an electron transition from valence band in a
conductivity band (i.e. the EHP creation) are essential. As well
as in previous works, the following model will be used. Vectors ${
\bf p}_{cv} $ for two degenerated valence bands $v=I $ and $v=II
$ are
\begin{eqnarray}
\label{5}{\bf p}_{cvI} ={p_{cv}\over\sqrt{2}}({\bf e}_{x} - i{\bf
e}_{ y}), \nonumber \\
{ \bf p} _{ cvII} ={p_{cv}\over\sqrt{2}}({\bf e}_{x} +i{\bf
e}_{y}),
\end{eqnarray}
where ${\bf e}_{x} $ and ${\bf e}_{y} $ are the unite vectors
along $x $ and $y $ axis, $ p _{ cv} $ is the real value. This
model corresponds to heavy holes in a semiconductor with the
 zinc blend structure, if the $z $ axis is directed along
a 4-th order axis [23,24]. If to use circular polarization
vectors of a stimulating light
\begin{equation}
\label{6}{ \bf e} _{ \ell} ={ 1\over\sqrt{ 2}} ({\bf e} _{ x} \pm
i{ \bf e} _{ y}),
\end{equation}
a property of preservation of a polarization vector
\begin{equation}
\label{7} \sum _{ v=I, II}{ \bf p} _{ cv} ^ * ({\bf e} _{ \ell}{
\bf p} _{ cv}) = \sum _{ v=I, II}{ \bf p} _{ cv} ({\bf e} _{
\ell}{ \bf p} _{ cv} ^ *) ={ \bf e} _{ \ell} p^2 _{ cv}
\end{equation}
is preserved. Thus, neither EHP wave functions, nor energy levels
depend on band numbers I or II.

  A class of a wave function $F_\rho ({\bf r}) $ at ${
\bf r} _e ={ \bf r} _h ={ \bf r} $ is essential in the theory in
the effective mass approximation (${ \bf r} _e ({\bf r} _h) $ is
the electron (hole) radius - vector) [22]. We suppose that
 it is possible to write down
\begin{equation}
\label{8} F_\rho ({\bf r}) =Q_\pi ({\bf r} _ \perp) \phi_\chi (z).
\end{equation}

The separation of variables is possible, if the Coulomb
interaction poorly influences on the movement of these particles
in  $xy $ plane. It occurs, if the quantizing magnetic field is
directed along $z $ axis and the condition
$$ a _{ exs} ^2\gg a_H^2, \eqno (8a) $$ is carried out,
where
$$ a _{ exs} = \hbar^2\varepsilon_0 / \mu e^2 $$
is the radius of a Wannier-Mott exciton in absence of a magnetic
field, $ \varepsilon_0 $ is the static permittivity, $ \mu $ is
the reduced effective mass, $a_H=\sqrt{c\hbar / | e|H} $ is the
magnetic length.

For $GaAs $ it is obtained (the parameters from [26] are used):
$$ a _{ exs} =146\AA, \quad a_H ^{ res} =57.2\AA, \eqno (8b) ,$$
where $a_H ^{ res} $ corresponds to magnetic field $H _{ res} $,
which is obtained from (1) at $j=1 $ for a magnetopolaron,
containing an electron. According to (8b) we obtain
$$ (a_H ^{ res} /a _{ res}) ^2\cong 0.154, $$
i.e. the condition (8 ) is carried out. Influence of the Coulomb
interaction on the movement of particles in $xy $ plane was
considered in [27].

In case of free EHPs (without an account of Coulomb forces )
\begin{equation}
 \label{9}\phi_\chi(z)=\varphi_{\ell_e}^e(z)\varphi_{\ell_h}^h(z),
\end{equation}
where $\varphi_\ell ^{ e (h)} (z) $ is the electron (hole) wave
function corresponding to a size quantization quantum number $l
$.  For QWs of finite depth functions $ \varphi_\ell ^{ e (h)} (z)
$ are given, for example, in [25]. For infinitely deep QW (when
there is no any tunnel penetration of electrons and holes in a
barrier)
$$\varphi_{\ell_e}^e(z)\,=\,\varphi_{\ell_h}^h(z)\,=\,\varphi_\ell(z),
~~~~~\ell=1,2,....$$
\begin{equation}
\label{10} \varphi_\ell(z)=\cases {\sqrt{2\over d}\quad sin({\pi
\ell z\over d} +{\pi \ell \over 2}), & $-{d\over 2}\leq z\leq
{d\over 2}$, \cr 0 & $z\leq -{d\over 2},\,z\geq {d\over 2}.$\cr}
\end{equation}

In order to neglect by Coulomb forces at the description
 of particles movements along  $z $ axis, the performance of the
condition
$$ a _{ exs} > d \eqno (10a) $$
is required. Whether the conditions (4) and (10 ) are compatible?
For example, for $GaAs $ the energy gap $ \hbar \omega_g\cong
1.85 eV $, the frequency of stimulating light owes to exceed this
value. The module $ \kappa_g =\omega_g\nu/c $ of a light wave
vector corresponds to the frequency $ \omega_g $, where $ \nu $
is the refraction index, $c $ is the light velocity in vacuum. For
$GaAs $ we have $ \kappa_g=3.16.10^5 cm ^{ -1} $, i. e. if $d=a
_{ exs} $, then $ \kappa_g a _{ exs} =0.46 $, what is comparable
to unit. Thus, for $GaAs $ it is possible to neglect  by Coulomb
forces at movement along  $z $ axis  to combine with condition $
\kappa d\geq 1 $ for " wide QWs " only with  some stretch. But
even if there are essential deviations from (9) we obtain the
same qualitative results for frequency dependence of light
absorption and reflection using unknown function $ \phi_\chi (z)
$ (see below).

We write down the stimulating electrical field extending along $z
$ axis as
\begin{equation}
\label{11}{ \bf E} _{ 0} (z, t) ={ {\bf e} _ \ell \over 2\pi}
\int _{ -\infty} ^{ \infty} d\omega e ^{ -i\omega t}{ \cal E} _0
(z, \omega) + c.c.,
\end{equation}
where $ \nu $ is the refraction index (identical inside  and
outside of a QW),
\begin{equation}
\label{12}{ \cal E} _{ 0} (z, \omega) =2\pi \, E _{ 0} e ^{
i\kappa z}{ \cal D} _{ 0} (\omega), ~~\kappa =\omega\nu/c.
\end{equation}
 ${ \cal D} _{ 0} (\omega) $ may correspond to a light pulse
 of any form [22] and
 at excitation  by monochromatic light with frequency
$ \omega _{ \ell} $ looks like
\begin{equation}
\label{13}{ \cal D}_0(\omega)=\delta(\omega-\omega_{\ell}).
\end{equation}

Let us spread out a true field ${ \bf E} (z, t) $ in the Fourier
integral
\begin{equation}
\label{14}{ \bf E} (z, t) ={ {\bf e} _ \ell\over 2\pi} \int _{
-\infty} ^{ \infty} d\omega e ^{ -i\omega t}{ \cal E} (z, \omega)
+ c.c..
\end{equation}
In [22] the equation (see also [19])
\begin{eqnarray}
 \label{15}&{\cal E}&(z, \omega) = -{ i\over
 2}\sum_\rho\gamma_{r\pi}\int_{-d/2}^{d/2}dz'\phi_\chi(z'){\cal
E} (z ', \omega) \nonumber \\
 &\times&\left\{e^{i\kappa z} \int _{ -d/2} ^zdz'' e ^{ -i\kappa
z ''}\phi_\chi (z'')\right.\nonumber \\
 &+&\left.e^{-i\kappa z} \int_z ^{ d/2} dz'' e ^{ i\kappa z''}
\phi_\chi (z'')\right \}\nonumber \\
 &\times&\{(\omega-\omega_\rho+i\gamma_\rho/2)^{-1}
 +(\omega+\omega_\rho+i\gamma_\rho/2)^{-1}\}\nonumber\\ &+&{\cal
E} _0 (z, \omega),
\end{eqnarray}
for the Fourier-component ${ \cal E} (z, \omega) $ is obtained,
where $ \kappa =\omega\nu/c $, the sum on indexes $ \rho $ means
summation on  excitation levels in a QW. A set of indexes $ \rho
$ is separable on two groups
\begin{equation}
\label{16} \rho\rightarrow\pi, \chi,
\end{equation}
concerning, according to (8), to cross and longitudinal wave
functions. In case of free EHPs the set $ \chi $ contains two
indexes: $ \ell_e $ and $ \ell_h $.

In a quantizing magnetic field (far from  magnetophonon resonance)
the set $ \pi $ includes only one index
\begin{equation}
\label{17} n_e=n_h=n,
\end{equation}
where $n_e (n_h) $ is the Landau quantum number concerning to
electron (hole). Close  to the magnetophonon resonance a level
with an index $n $ splits in two levels with indexes $p=a $ and
$p=b $ [25,21].

The equality (17) is caused by the law of preservation of quantum
number $n $ at EHP creation in case of normal light incidence on
QW surface.

Far away from the magnetophonon resonance energy levels in a
quantizing magnetic field are
\begin{equation}
\label{18}
 \omega_\rho=\omega_g+\varepsilon_\chi/\hbar+\Omega_{\mu H}(n+1/2),
\end{equation}
\begin{equation}
\label{19} \Omega _{ \mu H} ={ |e|H\over\mu c},
\end{equation}
$H $ is the magnetic field, $ \mu=m_em_h / (m_e+m_h). $
 In case of free EHPs
\begin{equation}
\label{20}
 \varepsilon_\chi=\varepsilon_{\ell_e}^e+\varepsilon_{\ell_h}^h,
\end{equation}
where $\varepsilon_{\ell_e}^e(\varepsilon_{\ell_h}^h)$ is the size
quantized electron (hole) energy level and in approximation of
infinitely deep QW (which is used at obtaining of (15))
\begin{equation}
\label{21} \varepsilon_\ell^{e(h)}={\hbar^2\pi^2\ell^2\over 2m _{
e (h)} d^2}.
\end{equation}
Close to the magnetophonon resonance for a polaron  A (Fig.1),
according, for example, to [21], we have
\begin{equation}
 \label{22}\omega_\rho=\omega_g+\varepsilon_\chi/\hbar+(3/2)\Omega_h+E_p/\hbar,
\end{equation}
\begin{equation}
\label{23}
 E_p=\hbar\Omega_e+\hbar\omega_{LO}/2\pm\sqrt{(\lambda/2)^2+A^2},
\end{equation}
 $$\lambda=\hbar(\Omega_e-\omega_{LO}),~~A=\Delta E/2, $$
$ \Delta E $ is the polaron splitting in the exact resonance, when
$ \lambda=0 $. The top sign in (23) corresponds to the top polaron
 level $p=a $, bottom - to the bottom polaron level $p=b $.
\begin{figure}
\rotatebox{90}{\includegraphics[height=\columnwidth]{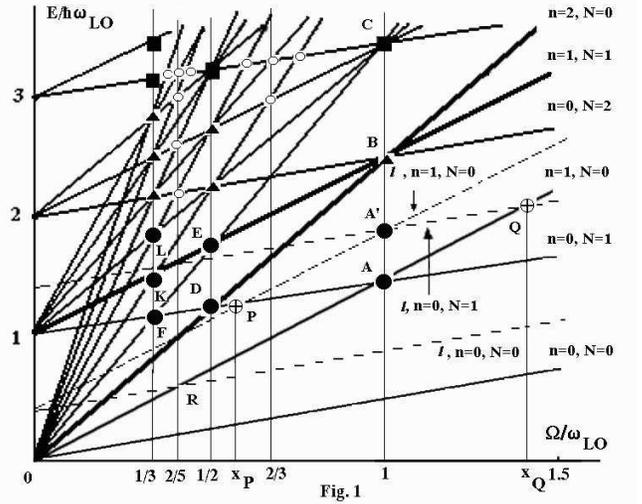}}
\caption[*]{\label{Fig1a.eps}Energy levels of of electron
(hole)-phonon system as
 functions of a magnetic field. Polaron states correspond to
crossing points of lines.
 Black circles are the twofold polarons,
triangles are threefold polarons, squares are fourfold polarons.
Empty circles are weak polarons. $ \Omega $ is the cyclotron
frequency, $ \omega _{ LO} $ is the $LO$-phonon frequency, $E $ is
the energy counted from the size quantized energy  $
\varepsilon_\ell $. $n $ is the Landau quantum number, $N $ is the
number of $LO $ - phonons, $ \ell, \ell ' $ are the size
quantization quantum numbers. }
\end{figure}

The values $ \gamma _{ r\pi} $ represent some factors which are
included in expressions for radiative broadenings of electronic
excitations (see (24)). In [21] these broadenings $ \tilde{
\gamma} _{r\rho} $ are calculated  in the distance and near the
magnetophonon resonance. At ${ \bf{ \cal K}} _ \perp=0 $ (where
${ \bf{ \cal K}} _ \perp$ is the cross component of a
quasi-momentum  of an electronic excitation) we have
\begin{equation}
 \label{24}\tilde{\gamma}_{r\rho}=\gamma_{r\pi}|R_\chi(\omega_\rho\nu/c)|^2,
\end{equation}
where
\begin{equation}
 \label{25}R_\chi(\kappa)=\int_{-d/2}^{d/2}dze^{-i\kappa
z} \phi_\chi (z).
\end{equation}
Far from  the magnetophonon resonance
\begin{equation}
\label{26} \gamma _{ r\pi} =2{ e^2\over\hbar c\nu}{ p _{ cv}
^2\over
 m_0}{\Omega_0\over\hbar\omega_g},~~\Omega_0={|e|H\over m_0c},
\end{equation}
i.e. it does not depend on an index $ \pi=n $. Near the resonance
for an excitation, consisting of the polaron $A $ (see Fig.1) and
a hole with $n=1 $
\begin{equation}
\label{27} \gamma _{ rp} =2{ e^2\over\hbar c\nu}{ p _{ cv}
^2\over m_0}{ \Omega_0\over\hbar\omega_g} Q _{ 0p},
\end{equation}
where
\begin{equation}
\label{28} Q _{ 0p} ={ 1\over
 2}\left(1\pm{\lambda\over\sqrt{\lambda^2+4A^2}}\right),
\end{equation}
and the top sign corresponds to the term $p=a $, bottom - to the
term $p=b $. Precisely in the resonance $ \lambda=0 $, and $Q _{
0p} $ is identical for terms $p=a $ and $p=b $ and is equal 1/2.
The factor $Q _{ 0p} $ depends strongly from the value $ \lambda $
deviation of the cyclotron frequency from the resonant value $
\Omega_e $  (see Fig.3 in [21]).

 In expressions (26) and (27) the approximation
 $ \omega_\rho\simeq\omega_g $ is used, what corresponds to the  effective
 mass method.

 At last, the values $ \gamma_\rho $ are  non-radiative
 broadenings of excitations with an index $ \rho $. In [21]
 their estimations from below for terms $p=a $ and $p=b $ are given.
\begin{figure}
\includegraphics[]{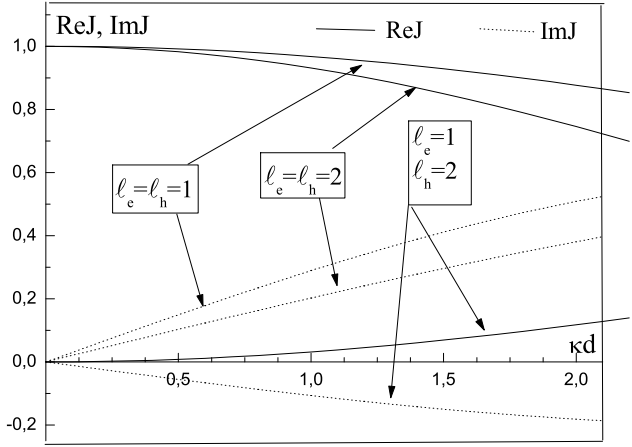}
\caption[*]{\label{Fig2.eps}Dependence of radiative damping $
\tilde{ \gamma} _r $ (continuous lines) and radiative shift $
\Delta $ of EHP energy (dashed lines) on a QW width $d $. On an
axis of ordinates  $Re J =\tilde{ \gamma} _r/\gamma_r $ and $Im
J=2\Delta/\gamma_r $ are shown, $J $ is calculated in (65). $ \ell
_{ e (h)} $ is the size quantization quantum number of electron
(hole), $ \kappa $ is the module of a light wave vector.}
\end{figure}

\section{Calculation of electric fields in case of two exited
energy levels}

 Let us limit the sum in (15) on numbers of exited levels by two terms:
 $i=1,2 $. It is possible, if two levels 1 and 2
 are located closely to each other and other levels are far from them- on
 the distance $ \Delta\omega $. Thus,
\begin{equation}
\label{29}
 \gamma_{1(2)}<<|\Delta\omega|,~~\gamma_{r1(2)}<<|\Delta\omega|.
\end{equation}
The level 1 is characterized by indexes $ \pi_1, \chi_1 $, level
 2 - by indexes $ \pi_2, \chi_2 $. Having entered new designations, we
rewrite (15) for a case of two levels as:
\begin{equation}
\label{30}{ \cal E} (z, \omega) ={ \cal E} _0 (z, \omega) +
\Delta{ \cal E} (z, \omega),
\end{equation}
\begin{eqnarray}
\label{31} \Delta{ \cal E} (z, \omega)&=&-{i\over
 2}\{\gamma_{r1}M_1(\omega)F_1(z)L_1(\omega)\nonumber\\
 &+&\gamma_{r2}M_2(\omega)F_2(z)L_2(\omega)\},
\end{eqnarray}
where $ \Delta{ \cal E} (z, \omega) $ is the Fourier-component of
a field, which we name "induced" as against components of an
exciting field ${ \cal E} _0 (z, \omega) $,
$$\gamma _{ r \, i} = \gamma _{ r\pi_i}, \quad \phi_i =\phi _{ \chi_i},
\quad \rho\to i, $$
\begin{equation}
\label{32}
 M_i(\omega)=\int_{-d/2}^{d/2}dz\phi_i(z){\cal E} (z, \omega),
\end{equation}
\begin{eqnarray}
\label{33} F_i(z)&=&e^{i\kappa z} \int _{ -d/2} ^{ z} dz '
e ^{ -i\kappa z '} \phi_i (z ') \nonumber \\
 &+&e^{-i\kappa z} \int ^{ d/2} _{ z} dz ' e ^{ i\kappa
z '} \phi_i (z '),
\end{eqnarray}
\begin{eqnarray}
\label{34}
 L_i(z)&=&(\omega-\omega_i+i\gamma_i/2)^{-1}\nonumber\\
 &+&(\omega+\omega_i+i\gamma_i/2)^{-1},~~ i=1,2.
\end{eqnarray}
The equation (31) describes electric fields at any  $ z $, i.e.
to the left of QW  at $z\leq -d/2 $, inside QW at $ -d/2\leq z
\leq d/2 $ and to the right of QW at $z\geq d/2 $.  It follows
from (31) and (33) that to the left of QW the component $ \Delta{
\cal E} _{ left} (z, \omega) $ is proportional to $\exp (-i\kappa
z) $, what corresponds to reflected light; to the right of QW the
component $ \Delta{ \cal E} (z, \omega) $ is proportional to
$\exp (i\kappa z) $, what corresponds to wave extending from left
to right. Inside QW the solution is more complicated.

The equations (30) - (31) may be solved by a method of
iterations, if the interaction of a stimulating field with an
electronic system is weak. Further we show that
 it is possible to be limited by a lowest approximation under the
condition
\begin{equation}
\label{35} \gamma _{ r1 (2)} < < \gamma _{ 1 (2)}.
\end{equation}
But we solve the equations (30) - (31) precisely, calculating
factors $M_1 (\omega) $ and $M_2 (\omega) $. For this purpose we
multiply  (30) consecutively on $ \phi_1 (z) $ and $ \phi_2 (z) $
and integrate on $z $ from $ -d/2 $ up to $d/2 $. We obtain the
system of two equations
\begin{eqnarray}
\label{36}
a _{ 11} M_1+a _{ 12} M_2=C_1, \nonumber \\
a _{ 21} M_1+a _{ 22} M_2=C_2,
\end{eqnarray}
where
$$ a _{ 11} =1 +{ i\over 2} \gamma _{ r1} L_1J _{ 11}, $$
$$ a _{ 12} ={ i\over 2} \gamma _{ r2} L_2J _{ 12}, $$
$$ a _{ 21} ={ i\over 2} \gamma _{ r1} L_1J _{ 21}, $$
$$ a _{ 22} =1 +{ i\over 2} \gamma _{ r2} L_2J _{ 22}, $$
\begin{equation}
\label{37} C_i =\int _{ -d/2} ^{ d/2} \, dz \,{ \cal E} _0 (z,
\omega) \, \phi_i (z),
\end{equation}
\begin{equation}
\label{38}
 J_{i\,i'}=\int_{-d/2}^{d/2}dz\phi_i(z)F_{i'}e.
\end{equation}
Solving the equation system (36) we obtain
\begin{equation}
\label{39} M_1 ={ C_1a _{ 22} -C_2a _{ 12} \over
 a_{11}a_{22}-a_{12}a_{21}},~~M_2={C_2a_{11}-C_1a_{21}\over
a _{ 11} a _{ 22} -a _{ 12} a _{ 21}}.
\end{equation}
Let us note the important properties of factors $C_i $ and $a _{ i
\, i '} $. Having substituted (12) in (37), we obtain
\begin{equation}
\label{40} C_i=2\pi E_0{ \cal D} _0 (\omega){ \cal R} _{ \chi_i} ^
* (\kappa),
\end{equation}
where
$$ R_i (\kappa) =R _{ \chi_i} (\kappa) =
\int _{ -d/2} ^{ d/2} dz ~~ exp (-i\kappa z) ~ \phi_i (z). $$
Having substituted (33) in (38), we have
\begin{eqnarray}
\label{41}
 J_{i\,i'}&=&
 \int_{-d/2}^{d/2}dz\phi_{\chi_i}(z)\left\{e^{i\kappa z
} \int _{ -d/2} ^zdz ' e ^{ -i\kappa z }
\phi_{\chi_{i'}}(z')\right.\nonumber\\&+& \left.e ^{ -i\kappa z}
\int_z ^{ d/2} dz ' e ^{ i\kappa z ' } \phi _{ \chi _{ i '}} (z ')
\right \},
\end{eqnarray}
what is easy to transform to
\begin{eqnarray}
 \label{42}J_{i\,i'}&=&\int_{-d/2}^{d/2}dz e ^{ i\kappa z
} \left \{ \phi _{ \chi_i (z)} \int _{ -d/2} ^zdz ' e ^{ -i\kappa
z}
\phi _{ \chi _{ i '}} (z ') \right.\nonumber \\
 &+& \left.\phi _{ \chi _{ i '}} (z) \int _{ -d/2} ^zdz ' e ^{ -i\kappa
z '} \phi _{ \chi_i} (z ') \right \}.
\end{eqnarray}
It follows from (41) and (42)
$$ J _{ i \, i '} =J _{ i ' \, i}, $$
\begin{eqnarray}
\label{43} Re J_{i\,i'}&={1\over 2} \{ {\cal R} _{ \chi_i} ^ *
(\kappa){ \cal R} _{ \chi _{ i '}} (\kappa) +{ \cal R} _{ \chi_i}
(\kappa){ \cal R} _{ \chi _{ i '}} ^ * (\kappa) \},
\end{eqnarray}
in particular,
\begin{equation}
\label{44} Re J _{ i \, i} = |{ \cal R} _{ \chi_i} (\kappa) | ^2.
\end{equation}
We enter also a designation
\begin{equation}
\label{45} q_{i\,i'}=ImJ_{i\,i'},~~q_{i\,i'}(\kappa=0)=0.
\end{equation}
Thus, having substituted (39) in (31) and using the formula
\begin{equation}
\label{46} \Delta{ \bf E} (z, t) ={ {\bf e} _l\over 2\pi} \int _{
-\infty} ^{ \infty} d\omega e ^{ -i\omega t} \Delta{ \cal E} (z,
\omega) +c.c.,
\end{equation}
we basically have solved the task about calculation of induced
fields in case of two exited energy levels in a wide QW. Further
we consider some special cases.
\begin{figure}
\includegraphics[]{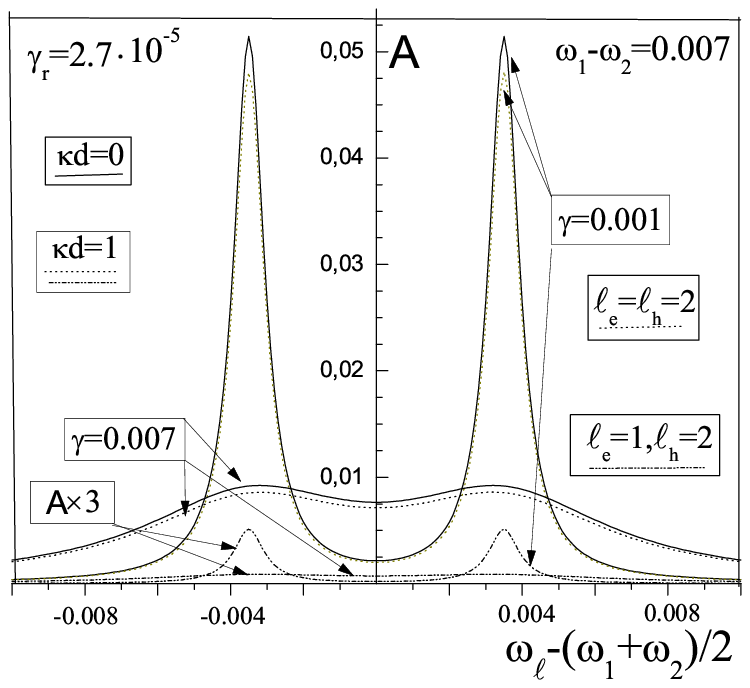}
\caption[*]{\label{Fig3a.eps}Dimensionless light  reflection ${
\cal R} $  as function of light frequency $ \omega_\ell $ in case
of two excitation energy levels
 in a wide QW under condition $ \gamma_r\ll\gamma $.
Continuous and dashed lines are the permitted transitions,
dot-and-dash lines are the forbidden transitions. $ \gamma_r
(\gamma) $ is the radiative (non-radiative) broadening of an
exited state.}
\end{figure}

\section{Interaction with one level}

If to put
\begin{equation}
\label{47} \gamma _{ r2} =0,
\end{equation}
it follows from (31) that it remains an interaction of light only
with one level. Let us enter designations
\begin{eqnarray}
\label{48} \gamma _{ r1} = \gamma_r,~~ \omega_1 =\omega_0, ~~
L_1 (\omega) =L (\omega), \nonumber \\
{ \cal R} _1 (\kappa) ={ \cal R} (\kappa), ~~ q_1 (\kappa) =q
(\kappa)
\end{eqnarray}
and obtain following expressions for fields at the left and to
the right of QW:
\begin{eqnarray}
\label{49} \Delta{ \bf E}_{left}(z,t)&=&-{i\over 2}{ \bf e} _
\ell E_0\int _{ -\infty} ^{ \infty} dt e ^{ -i\omega t-i\kappa
z+i\alpha} \nonumber \\
 &\times&{{\tilde \gamma} _r (\omega) L (\omega){ \cal D} _0 (\omega) \over
1 + (i{ \tilde \gamma}_r(\omega)/2-\Delta(\omega))L(\omega)}+c.c.,
\end{eqnarray}
$$\Delta{ \bf E} _{ right} (z, t) = -{ i\over 2}{ \bf
e} _ \ell E_0\int _{ -\infty} ^{ \infty} dt e ^{ -i\omega
t+i\kappa z} $$
$$\times{ {\tilde \gamma} _r (\omega) L (\omega){ \cal
D} _0 (\omega) \over 1 + (i{ \tilde
 \gamma}_r(\omega)/2-\Delta(\omega))L(\omega)}+c.c.,\eqno(49a)$$
where
\begin{eqnarray}
\label{50}{ \tilde \gamma} _r (\omega) = \gamma_r |{ \cal R}
(\kappa) | ^2, ~~
\Delta (\omega) = \gamma_r q (\kappa) /2, \nonumber \\
e ^{ i\alpha} ={ {\cal R} ^ * (\kappa) \over{ \cal R} (\kappa)}.
\end{eqnarray}
If to use (9) and the function (10), it is possible to show the
following. If the size quantization quantum numbers   $ \ell_e $
and $ \ell_h $ are of identical parity, ${ \cal R} ^ * (\kappa) ={
\cal R} (\kappa) $ and $e ^{ i\alpha} =1 $, and if they are
different, ${ \cal R} ^
* (\kappa) = -{ \cal R} (\kappa) $ and $e ^{ i\alpha} = -1 $.
\begin{figure}
\includegraphics[]{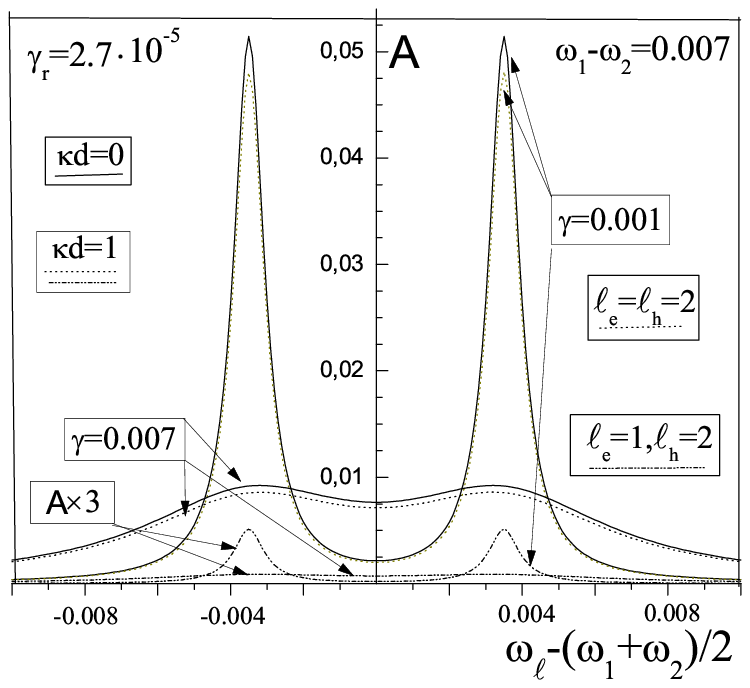}
\caption[*]{\label{Fig3b.eps}Dimensionless light absorption ${
\cal A} $ as function of light frequency $ \omega_\ell $ in case
of two excitation energy levels
 in a wide QW under condition $ \gamma_r\ll\gamma $.
Continuous and dashed lines are the permitted transitions,
dot-and-dash lines are the forbidden transitions. $ \gamma_r
(\gamma) $ is the radiative (non-radiative) broadening of an
exited state.}
\end{figure}

If to reject in expression
 $$L(\omega)=(\omega-\omega_0+i\gamma/2)^{-1}+(\omega+\omega_0+i\gamma/2)^{-1}$$
 the non-resonant term $ (\omega +\omega_0+i\gamma/2)
^{ -1} $, then from (49) and (49a) we obtain
\begin{eqnarray}
\label{51} \Delta{ \bf E} _{ left} (z, t)&=&-{i\over 2}{ \bf e} _
\ell E_0\int _{ -\infty} ^{ \infty} d\omega e ^{ -i\omega
t-i\kappa
z+i\alpha} \nonumber \\
 &\times&{{\tilde \gamma} _r (\omega){ \cal D} _0 (\omega) \over
\omega- (\omega_0 +\Delta) +i ({\tilde \gamma} _r +\gamma) /2},
\end{eqnarray}
\begin{eqnarray}
\label{52} \Delta{ \bf E} _{ right} (z, t)&=&-{i\over 2}{ \bf e}
_ \ell E_0\int _{ -\infty} ^{ \infty} d\omega e ^{ -i\omega
t+i\kappa
z} \nonumber \\
 &\times&{{\tilde \gamma} _r (\omega){ \cal D} _0 (\omega) \over
\omega- (\omega_0 +\Delta) +i ({\tilde \gamma} _r +\gamma) /2}.
\end{eqnarray}
Comparing the first equality from (50) to the expression (24), we
find that ${ \tilde \gamma} _r (\omega) $ is the radiative
broadening of the exited state with energy  $ \omega $. The value
$ \Delta (\omega) $ means a certain shift of the exited level in
QW caused by interaction with light.

If in [19,20] to put $ \nu_1 =\nu $ (where $ \nu_1 (\nu) $ is the
refraction index of a QW (barrier) substance), we obtain the
results (51) and (52) .\footnote{ In [19] in formulas (47) and
(48) contain misprints. Instead of $ \gamma_r $ it follows $
\tilde{ \gamma} _rexp (-i\kappa d/2) $.}

In case of narrow QWs ($ \kappa d < < 1 $) we obtain
\begin{equation}
\label{53} \tilde{ \gamma} _r \simeq \gamma_r \int _{ -d/2} ^{
d/2} dz \phi (z), ~~ exp (i\alpha) \simeq 1, ~~ \Delta (\omega)
=0,
\end{equation}
and for free electrons and holes (when (9) is carried out) $
\tilde{ \gamma} _r = \gamma_r \delta _{ \ell_e, \ell_h}, $
the expressions (51) and (52) pass in used, for example, in [16].

Let us emphasize that in case of wide QWs there is a dependence of
induced fields on a QW width $d $ (through values ${ \tilde
\gamma} _r (\omega) $ and $ \Delta (\omega) $), and also an
opportunity of interaction of light with EHPs, which have $
\ell_e\neq \ell_h $, appears.

\section{Two energy levels in a narrow QW}

In narrow QWs ($ \kappa d < < 1 $) under the condition (9) we
obtain from (37)
\begin{equation}
\label{54} C_i=2\pi E_0{ \cal D} _0\delta _{ \ell _{ e \, i}, \ell
_{ h, \, i}},
\end{equation}
and it  follows from (43)
\begin{equation}
\label{55} J _{ i \, i} = \delta _{ \ell _{ e \, i}, \ell _{ h,
\, i}}, ~~ J _{ 12} =J _{ 21} = \delta _{ \ell _{ e1}, \ell _{
h1}} \delta _{ \ell _{ e2}, \ell _{ h2}},
\end{equation}
i.e. light interacts with two levels, which size quantization
quantum numbers of electron and hole are identical, i.e.
$$\ell _{ e1} = \ell _{ h1}, ~~\ell _{ e2} = \ell _{ h2}. $$
We obtain also
\begin{eqnarray}
 \label{56}M_1&=&M_2\nonumber\\
 &=&{2\pi E_0{ \cal D} _0 (\omega) \over
 1+(i/2)[\gamma_{r1}L_1(\omega)+\gamma_{r2}L_2(\omega)]}.
\end{eqnarray}
Having substituted (56) in (31), we find
\begin{eqnarray}
\label{57} \Delta{ \cal E}(z,\omega)&=&-i\pi E_0{ \cal
D} _0 (\omega) \nonumber \\
 &\times&{\gamma_{r1}L_1(\omega)F_1(z)+\gamma_{r2}L_2(\omega)F_2(z)\over
 1+(i/2)[\gamma_{r1}L_1(\omega)+\gamma_{r2}L_2(\omega)]}.
\end{eqnarray}
Having substituted (57) in (46), we determine the induced
electric fields at the left and to the right of QW:
\begin{equation}
\label{58} \Delta{ \bf E} _{ left} (z, t) = {e} _lE_0\int _{
-\infty} ^ \infty d\omega e ^{ -i\omega t-i\kappa z}{ \cal D}
(\omega) +c.c.,
\end{equation}
$$\Delta{ \bf E}_{right}(z,t)={e}_lE_0\int_{-\infty}^\infty
d\omega e ^{ -i\omega t+i\kappa z}{ \cal D} (\omega) +c.c.,\eqno
(58a) $$
where
\begin{equation}
\label{59}{ \cal D} (\omega) = -{ 4\pi\chi (\omega){ \cal D} _0
(\omega) \over 1+4\pi\chi (\omega)},
\end{equation}
\begin{equation}
\label{60} \chi (\omega) ={ i\over 8\pi }
[\gamma_{r1}L_1(\omega)+\gamma_{r2}L_2(\omega)].
\end{equation}
For an electric field inside a QW we obtain the result
\begin{eqnarray}
\label{61} \Delta{ \bf E} _{ QW} (z, t)&=&-{i\over 2}{ \bf e}
_lE_0\int _{ -\infty} ^ \infty d\omega e ^{ -i\omega t}{ \cal
D} _0 (\omega) \nonumber \\
 &\times&\{1+(i/2)[\gamma_{r1}L_1(\omega)+\gamma_{r2}L_2(\omega)]\}^{-1}\nonumber\\
 &\times&\left\{\gamma_{r1}L_1(\omega)\left[e^{i\kappa
z} \int _{ -d/2} ^zdz '\phi_1 (z ') \right.\right.\nonumber \\
 &+&\left.e^{-i\kappa
z} \int ^{ d/2} _zdz '\phi_1 (z ') \right] \nonumber \\
 &+&\gamma_{r2}L_2(\omega)\left[e^{i\kappa
z} \int _{ -d/2} ^zdz '\phi_2 (z ') \right.\nonumber \\
 &+&\left.\left.e^{-i\kappa
z} \int ^{ d/2} _zdz '\phi_2 (z ') \right] \right \}.
\end{eqnarray}
Obviously, formulas (58), (58a) and (61) may be generalized on any
number of exited levels in narrow QWs (see, for example, [18]).
Expressions (58), (58a) were used in [17,21].

\section{Two levels in a wide QW. Electric fields.}

Let us consider a special case, when functions $ \phi _{ \chi_i}
(z) $ for two levels coincide, i.e.
\begin{equation}
 \label{62}\phi_{\chi_1}(z)=\phi_{\chi_2}(z)=\phi(z).
\end{equation}
If (9) is carried out, (62) is right, in particular, at
\begin{equation}
\label{63} \ell _{ e1} = \ell _{ e2} =l _{ e}, ~~\ell _{ h1} =
\ell _{ h2} = \ell _{ h}.
\end{equation}
Let us admit that  only $ \gamma _{ r1} $ and $ \gamma _{ r2} $,
energy $ \hbar\omega_1 $ and $ \hbar\omega_2 $ (see [21]) and
quantum numbers $ \ell_e $ and $ \ell_h $ may differ each from
other. Two close located levels of a system consisting of an usual
magnetopolaron  and a hole concern to such cases.

With the help of (31) and (39) we obtain
\begin{eqnarray}
\label{64} \Delta{ \cal E}(z, \omega)&=&-i\pi E_0{\cal
D}_0(\omega)\nonumber\\
 &\times&{[\gamma_{r1}L_1(\omega)+\gamma_{r2}L_2(\omega)]F(z)\over
 1+{i\over 2
}[\gamma_{r1}L_1(\omega)+\gamma_{r2}L_2(\omega)]J(\kappa)},
\end{eqnarray}
$L_i (\omega) $ is determined in (34), $F (z) $ is determined in
(33),
\begin{eqnarray}
 \label{65}J(\kappa)&=&\int_{-d/2}^{d/2}dz\phi(z)\left\{e^{i\kappa
z} \int _{ -d/2} ^{ z} dz ' e ^{ -i\kappa
z '} \phi (z ') \right.\nonumber \\
 &+&\left.e^{-i\kappa z} \int ^{ d/2} _{ z} dz ' e ^{ i\kappa
z '} \phi (z ') \right \},
\end{eqnarray}
\begin{equation}
\label{66} J (\kappa) = |{ \cal R} (\kappa) | ^2+iq (\kappa).
\end{equation}
For induced electrical fields on the left and to the right of QW
with the help (46) and (64) we find:
\begin{equation}
\label{67} \Delta{ \bf E} _{ left} (z, t) ={ \bf e} _lE_0\int _{
-\infty} ^ \infty d\omega e ^{ -i\omega t-i\kappa z+i\alpha}
\tilde{ {\cal D}} (\omega) + c.c.,
\end{equation}
$$\Delta{ \bf E} _{ right} (z, t) ={ \bf e} _lE_0\int _{ -\infty} ^ \infty
d\omega e ^{ -i\omega t+i\kappa z} \tilde{ {\cal D}} (\omega) +
c.c., ~~~ (67a) $$
$$\tilde{ {\cal D}} (\omega) = -{ i\over 2}{ \cal D}_0(\omega)
[\tilde{\gamma}_{r1}(\omega)L_1(\omega)
 +\tilde{\gamma}_{r2}(\omega)L_2(\omega)]$$
 $$\times\{1-\Delta_1(\omega)L_1(\omega)-\Delta_2(\omega)L_2(\omega)$$
 $$+(i/2)[\tilde{\gamma}_{r1}(\omega)L_1(\omega)
 +\tilde{\gamma}_{r2}(\omega)L_2(\omega)]\}^{-1},$$
 where
\begin{eqnarray}
 \label{68}\tilde{\gamma}_{r1(2)}(\omega)&=&\gamma_{r1(2)}|{\cal R
} (\kappa) | ^2, \nonumber \\
 \Delta_{1(2)}(\omega)&=&{1\over
2} \gamma _{ r1 (2)} q (\kappa), \nonumber \\
 e^{i\alpha}&=&{{\cal R} ^ * (\kappa) \over{ \cal R} (\kappa)}.
\end{eqnarray}
We reject non-resonant contributions to $L_1 (\omega) $ and $L_2
(\omega) $ and obtain
\begin{eqnarray}
\label{69} \tilde{ \cal
 D}(\omega)&=&-iD_0(\omega)\nonumber\\
 &\times&\left({\bar{\gamma}_{r1}/2\over
 \omega-\Omega_1+iG_1/2}+{\bar{\gamma}_{r2}/2\over
\omega-\Omega_2+iG_2/2} \right),
\end{eqnarray}
where $o _{ 1 (2)} = \Omega _{ 1 (2)} -iG _{ 1 (2)} /2, $ ($
\Omega $ and $G $ are real by definition) satisfy to the equation
\begin{eqnarray}
 \label{70}&(&o-\omega_1+i\gamma_1/2)(o-\omega_2+i\gamma_2/2)\nonumber\\
 &-&\Delta_1(o-\omega_2+i\gamma_2/2)\nonumber\\&-&
 \Delta_2(o-\omega_1+i\gamma_1/2)+i(\tilde{\gamma}_{r1}/2)(o-\omega_2+i\gamma_2/2)\nonumber\\
 &+&i(\tilde{\gamma}_{r2}/2)(o-\omega_1+i\gamma_1/2)=0
\end{eqnarray}
and are equal
\begin{eqnarray}
 \label{71}(\Omega&-&iG/2)_{1(2)}={1\over
 2}\{\tilde{\omega}_1+\tilde{\omega}_2\nonumber\\
 &\pm&\sqrt{(\tilde{\omega}_1-\tilde{\omega}_2)^2
 +(2\Delta_1-i\tilde{\gamma}_{r1})(2\Delta_2-i\tilde{\gamma}_{r2})}\};
\end{eqnarray}
the sign plus concerns to the subscript 1, and the sign minus - to
the subscript 2,
\begin{equation}
 \label{72}\tilde{\omega}_{1(2)}=\omega_{1(2)}+\Delta_{1(2)}-i\Gamma_{1(2)}/2,
\end{equation}
\begin{equation}
 \label{73}\Gamma_{1(2)}=\tilde{\gamma}_{r1(2)}+\gamma_{1(2)}.
\end{equation}
The expression in  the root in (71), generally speaking, is
complex. Designations  are entered also
\begin{equation}
 \label{74}\bar{\gamma}_{r1}=\tilde{\gamma}_{r1}+\Delta\gamma,~~
 \bar{\gamma}_{r2}=\tilde{\gamma}_{r2}-\Delta\gamma,
\end{equation}
\begin{eqnarray}
 \label{75}\Delta\gamma={\tilde{\gamma}_{r1}[\Omega_2-\omega_2-i(G_2-\gamma_2)]
\over \Omega_1-\Omega_2+i(G_2-G_1)/2}\nonumber\\
 +{\tilde{\gamma}_{r2}[\Omega_1-\omega_1-i(G_1-\gamma_1)]\over
\Omega_1-\Omega_2+i (G_2-G_1) /2}.
\end{eqnarray}

Let us compare results (67) and (67a) with substitution (69) to
the formulas (51) and (52) for a case of one excitation level in a
wide QW. It is visible that levels with numbers 1 and 2 influence
each other, that results into a renormalization of  $ \omega_i
+\Delta_i, \, \, \Gamma_i $ and $ \gamma _{ ri} $: They are being
replaced accordingly by $ \Omega_i, \, \, G_i $ and $ \bar{
\gamma} _{ ri} $. In narrow QWs at $ \kappa d\ll 1\quad
\Delta_1\cong\Delta_2\cong0, \quad \exp (i\kappa d) =1, $ and at
performance (9), i.e. without the Coulomb forces account,
\begin{equation}
\label{76}
 \tilde{\gamma}_{r1}=\gamma_{r1}\delta_{\ell_e,\ell_h},\qquad
 \tilde{\gamma}_{r2}=\gamma_{r2}\delta_{\ell_e,\ell_h},\qquad
\end{equation}
and we obtain results of [21] for two levels in a narrow QW.

\section{Two energy levels in a wide QW. Light reflection and
absorption  at monochromatic irradiation}

Knowing expressions for electric fields on the left and to the
right of QW, let us calculate light reflection and absorption.
Let us introduce Umov-Pointing vectors   ${ \bf S} _{ left
(right)} $ on the left (to the right) from QW
\begin{equation}
\label{77}{ \bf S} _{ left} ={ \bf S} _0 +\Delta{ \bf S} _{ left},
\end{equation}
\begin{equation}
\label{78}{ \bf S} _{ right} ={ c\nu\over 4\pi} ({\bf E} _0
+\Delta{ \bf E } _{ right}) ^2{ \bf e} _z,
\end{equation}
\begin{equation}
\label{79} \Delta{ \bf S} _{ left} = - (\Delta{ \bf E } _{ left})
^2{ c\nu\over 4\pi}{ \bf e} _z,
\end{equation}
\begin{equation}
\label{80}{ \bf S} _0 ={ c\nu\over 4\pi} E_0^2{ \bf e} _z.
\end{equation}
Dimensionless  light reflection is defined as
\begin{equation}
\label{81}{ \cal R} ={ | \Delta{ \bf S} _{ left} | \over |{ \bf
S} _0 |}.
\end{equation}
Light absorption is defined as
\begin{equation}
\label{82}{ \cal A} ={ | \Delta{ \bf S} _{ left} -\Delta{ \bf S}
_{ right} | \over |{ \bf S} _0 |}.
\end{equation}
Light transmission is equal
\begin{equation}
\label{83}{ \cal T} ={ |{ \bf S} _{ right} | \over |{ \bf S} _0
|} = 1-{ \cal R} -{ \cal A}.
\end{equation}
Let us consider a case of monochromatic irradiation of a QW, when
(13) is carried out.
With the help of (67), (67a), (80). (81) and (82) we find
\begin{eqnarray}
\label{84}{ \cal R}&=&{1\over
 4Z}\{[\tilde{\gamma}_{r1}(\omega_l-\omega_2)+\tilde{\gamma}_{r2}(\omega_l-\omega_1)]^2
\nonumber \\
 &+&{1\over
 4}(\tilde{\gamma}_{r1}\gamma_2+\tilde{\gamma}_{r2}\gamma_1)^2\},
\end{eqnarray}
\begin{eqnarray}
\label{85}{ \cal A}&=&{1\over
 2Z}\{\tilde{\gamma}_{r1}\gamma_1[(\omega_l-\omega_2)^2+\gamma_2^2/4]\nonumber\\
 &+&\tilde{\gamma}_{r2}\gamma_2[(\omega_l-\omega_1)^2+\gamma_1^2/4]+(\Delta_1\tilde{\gamma}_{r2}-\Delta_2\tilde{\gamma}_{r1})
 \nonumber\\
 &\times&[(\omega_l-\omega_2)\gamma_1-
(\omega_l-\omega_1) \gamma_2]\},
\end{eqnarray}
where
\begin{eqnarray}
\label{86}
 \tilde{\gamma}_{r1(2)}=\tilde{\gamma}_{r1(2)}(\omega_l)\cong
 \tilde{\gamma}_{r1(2)}(\omega_g),\nonumber\\
\Delta _{ 1 (2)} = \Delta _{ 1 (2)} (\omega_l)
\cong \Delta _{ 1 (2)} (\omega_g), \nonumber \\
 Z=[(\omega_l-\Omega_1)^2+G_1^2/4][(\omega_l-\Omega_2)^2+G_2^2/4].
\end{eqnarray}
Using (71) - (73), we transform (86) to
\begin{eqnarray}
 \label{87}Z&=&\{(\omega_l-\omega_1)(\omega_l-\omega_2)-{1\over
 4}(\tilde{\gamma}_{r1}\gamma_2+\tilde{\gamma}_{r2}\gamma_1+\gamma_{1}\gamma_2)\nonumber\\
 &-&\Delta_1(\omega_l-\omega_2)-\Delta_2(\omega_l-\omega_1)\}^2\nonumber\\
 &+&{1\over
 4}\{(\omega_l-\omega_1)\Gamma_2+(\omega_l-\omega_2)\Gamma_1\nonumber\\&-&\Delta_1\gamma_2-\Delta_2\gamma_1
\} ^2.
\end{eqnarray}
Let us emphasize that expressions (84) and (85) for light
reflection and absorption, as well as expressions (67) and (67a)
for induced electric fields, are right under condition (8), but
performance of the condition (9) is not necessarily. The Coulomb
forces may be taken into account, therefore functions $ \phi _{
\chi_i} (z) $ will differ from (9). In narrow QWs at $ \kappa
d\ll 1 $ we obtain
\begin{equation}
\label{88} { \cal R} _ \chi = \int _{ -\infty} ^{ \infty} dz
\phi_\chi (z), \quad \Delta_\chi \simeq 0.
\end{equation}

Let us consider some extrem cases.

1. Under condition $ \kappa d < < 1 $, i.e. for narrow QWs, we use
(76). Then the expressions (84) and (85) pass in appropriate
formulas from [21]. Let us emphasize that in the limit $ \kappa d
< < 1 $ for free EHPs  only $ \ell_e =\ell_h $ is admitted.

2. Having put $ \gamma _{ r2} =0 $, we exclude interaction of
light with the level 2. Then we obtain results for one level
\begin{equation}
\label{89}{ \cal R} ={ \tilde{ \gamma} _{ r1} ^2\over
 4[(\omega_l-\omega_1-\Delta_1)^2+(\tilde{\gamma}_{r1}+\gamma_1)^2/4]},
\end{equation}
\begin{equation}
\label{90}{ \cal A} ={ \tilde{ \gamma} _{ r1} \gamma_1\over
 2[(\omega_l-\omega_1-\Delta_1)^2+(\tilde{\gamma}_{r1}+\gamma_1)^2/4]}.
\end{equation}

3. Let us consider an extreme case
\begin{equation}
\label{91} \tilde{ \gamma} _{ r1 (2)} < <
 \gamma_{1(2)},~~\Delta_{1(2)}<<\gamma_{1(2)},
\end{equation}
when the perturbation theory on light interaction with an
electronic system is applicable. From (84) and (85) we find
\begin{eqnarray}
\label{92}{\cal R} ={(\tilde{\gamma}_{ r1} /2)^2\over(
\omega_l-\omega_1)^2+(\gamma_1/2)^2}+{(\tilde{\gamma}_{r2}/2)^2\over
(\omega_l-\omega_2)^2+(\gamma_2/2)^2}\nonumber\\
 +{\tilde{\gamma}_{r1}\tilde{\gamma}_{r2}\over
 2}{(\omega_l-\omega_1)(\omega_l-\omega_2)+\gamma_1\gamma_2/4\over
[(\omega_l-\omega_1)^2+(\gamma_1/2)^2][(\omega_l-\omega_2)^2+(\gamma_2/2)^2]},
\end{eqnarray}
\begin{eqnarray}
\label{93}{ \cal A}\simeq{\tilde{\gamma}_{r1}\gamma_1/2\over (
\omega_l-\omega_1)^2+(\gamma_1/2)^2}+{\tilde{\gamma}_{r2}\gamma_2/2\over
(\omega_l-\omega_2) ^2 + (\gamma_2/2) ^2}.
\end{eqnarray}
Since
\begin{eqnarray}
\label{94} \tilde{\gamma}_{r1(2)}&=&\gamma_{r1(2)}|{\cal
R} (\kappa_l) | ^2, \nonumber \\
{ \cal R}(\kappa_l)&=
 &\int_{-\infty}^{\infty}dze^{-i\kappa_lz}\phi (z),
~~\kappa_l ={ \omega_l\nu\over c},
\end{eqnarray}
and  reflection (92) is square on values $ \tilde{ \gamma} _{ r}
$, in the RHS of (92) the factor $ |{ \cal R} (\kappa_l) | ^4 $
may be taken out of brackets. Similarly in (93) it is possible to
bear out brackets $ |{ \cal R} (\kappa_l) | ^2 $. In brackets  $
\tilde{ \gamma} _{ r1 (2)} $ should be replaced by $ \gamma _{ r1
(2)} $. Thus, dependence of ${ \cal R} $ and ${ \cal A} $ on QW
width is determined by factors $ |{ \cal R } (\kappa_l) | ^4 $
and $ |{ \cal R} (\kappa_l) | ^2 $, respectively.

For free EHPs in a case $ \kappa_l d\ge1 $ there appears an
opportunity of light interaction with EHPs for which $ \ell_e\ne
\ell_h $.

According to (93) light absorption  ${ \cal A} $ is the sum of
 contributions from levels 1 and 2. Each of these contributions
may be obtained from (90) at
 $\tilde{\gamma}_{r}<<\gamma,~~\Delta<<\gamma$.

According to (92) light reflection ${ \cal R} $ is square  on $
\tilde{ \gamma} _{ r1} $ and $ \tilde{ \gamma} _{ r2} $ and
consequently, except of contributions of separate levels, contains
 an interference contribution.

4. The following limiting case is opposite to previous and is
determined by conditions
\begin{equation}
\label{95} \gamma _{ r1 (2)} > >
 \gamma_{1(2)},~~\Delta_{1(2)}>>\gamma_{1(2)}.
\end{equation}
Let us assume in the RHS of (84) and (85)
\begin{equation}
\label{96} \gamma _{ 1} = \gamma _{ 2} =0,
\end{equation}
then ${ \cal A} =0 $,
\begin{eqnarray}
\label{94}{ \cal
 R}&=&[(\tilde{\gamma}_{r1}+\tilde{\gamma}_{r2})/2]^2(\omega_l-\Omega_0)^2\nonumber\\
 &\times&\{[(\omega_l-\omega_1)(\omega_l-\omega_2)\nonumber\\
 &-&
 \Delta_1(\omega_l-\omega_2)-\Delta_2(\omega_l-\omega_1)]^2\nonumber\\&+&[(\tilde{\gamma}_{r1}
 +\tilde{\gamma}_{r2})/2]^2(\omega_l-\Omega_0)^2\}^{-1},
\end{eqnarray}
where
\begin{equation}
 \label{98}\Omega_0={\omega_1\tilde{\gamma}_{r2}+\omega_2\tilde{\gamma}_{r1}\over
 \tilde{\gamma}_{r1}+\tilde{\gamma}_{r2}}.
\end{equation}
Having determined roots of the equation
 $$(\omega_l-\omega_1)(\omega_l-\omega_2)-\Delta_1(\omega_l-\omega_21)$$
$$-\Delta_2 (\omega_l-\omega_1) =0, $$
let us transform (97) to
\begin{eqnarray}
\label{99}{ \cal
 R}&=&[(\tilde{\gamma}_{r1}+\tilde{\gamma}_{r2})/2]^2(\omega_l-\Omega_0)^2\nonumber\\
 &\times&\{(\omega_l-\omega_{d1})^2(\omega_l-\omega_{d2})^2
 \nonumber\\&+&[(\tilde{\gamma}_{r1}
 +\tilde{\gamma}_{r2})/2]^2(\omega_l-\Omega_0)^2\}^{-1},
\end{eqnarray}
where
\begin{eqnarray}
 \label{100}\omega_{d1(2)}&=&{1\over
 2}\{\omega_1+\Delta_1+\omega_2+\Delta_2\nonumber\\
 &\pm&\sqrt{(\omega_1+\Delta_1-\omega_2-\Delta_2)^2+
4\Delta_1\Delta_2} \}.
\end{eqnarray}
It follows from (99) that ${ \cal R} =0 $ at $ \omega_\ell
=\Omega_0 $ , i.e. there is a point of total light transmission
through a QW. At $ \omega_\ell =\omega _{ d1} $ or $ \omega_\ell
=\omega _{ d2} $ ~~ ${ \cal R} =1 $, i.e. light is completely
reflected. Let us compare this result (99) with the result of [21]
for narrow QWs ($ \kappa d < < 1 $). We find that at transition to
a case of wide QWs values $ \tilde{ \gamma} _{ r1 (2)} $ begin to
depend on a QW width, the interaction with EHPs for which $
\ell_e\ne \ell_h $ is admitted and points of total reflection are
displaced (at $ \kappa d < < 1 $ these points are $ \omega_\ell
=\omega_1 $ and $ \omega_\ell =\omega_2 $).

But existence of one point of total transmission and two points
of total reflection in case of wide QWs are preserved (see Fig.7
below).

5. Following extreme case is the smallness of broadenings and
energy shifts in comparison to distance $ \omega_1-\omega_2 $
between energy levels, i.e.
\begin{eqnarray}
 \label{101}\tilde{\gamma}_{r1(2)}<<\omega_1-\omega_2,~~~
 \Delta_{1(2)}<<\omega_1-\omega_2,\nonumber\\
\gamma _{ 1 (2)} < < \omega_1-\omega_2,
\end{eqnarray}
thus, a ratio between non-radiative broadening  $ \gamma _{ 1 (2)}
$ and values $ \tilde{ \gamma} _{ r1 (2)} $ and $ \Delta _{ 1 (2)}
$ may be anyone.

Let frequency $ \omega_\ell $ is close to a resonance with a level
1, i.e. in addition to (101) the conditions are carried out
\begin{eqnarray}
 \label{102}\omega_\ell-\omega_1<<\omega_\ell-\omega_2,~~~
 \Gamma_{1(2)}<<\omega_\ell-\omega_2,\nonumber\\
\Delta _{ 1 (2)} < < \omega_\ell-\omega_2.
\end{eqnarray}
Then from (84) and (85) we obtain results (89) and (90), i.e. the
second level poorly influences ${ \cal R} $ and ${ \cal A} $.

6. At last, we consider a case of merging levels, when
\begin{eqnarray}
 \label{103}\omega_\ell=\omega_2=\omega_0,~~~
 \tilde{\gamma}_{r1}=\tilde{\gamma}_{r2}=\tilde{\gamma}_{r},\nonumber\\
\Delta_1 =\Delta_2 =\Delta, ~~~ \gamma_1 =\gamma_2 =\gamma.
\end{eqnarray}
From (71) we obtain
\begin{eqnarray}
\label{104} \Omega_1 =\omega_0+2\Delta, ~~~ \Omega_2 =\omega_0,
\nonumber \\ G_1=2\tilde{ \gamma} _{ r} + \gamma, ~~~ G_2 =\gamma,
\end{eqnarray}
Then it follows from (84) - (86)
\begin{equation}
\label{105}{ \cal
 R}={\tilde{\gamma}_{r}^2\over(\omega_\ell-\omega_0-2\Delta)^2+(2\tilde{\gamma}_{r}+\gamma)^2/4},
\end{equation}
\begin{equation}
\label{106}{ \cal
 A}={\tilde{\gamma}_{r}\gamma\over(\omega_\ell-\omega_0-2\Delta)^2
+ (2\tilde{ \gamma} _{ r} + \gamma) ^2/4}.
\end{equation}
Comparing obtained results with (89) and (90), we find that in
case of twice degenerated exited level  the formulas for one
non-degenerated level with the doubled values of $ \tilde{ \gamma}
_{ r} $ and $ \Delta $ are right.

\section{The classification of magnetopolarons}

In Fig.1 terms of an electron-phonon system in a QW, concerning
to a size quantization quantum number $ \ell $, are represented
by continuous lines. It is supposed that phonons, essential at
formation of magnetopolarons (confined or of interface), have one
frequency $ \omega _{ LO} $ without the account of dispersion. On
an abscissa axis the relation  $j ^{ -1} = \Omega _{ e (h)}
/\omega _{ LO} $ is shown, on an ordinates axis  the relation
$E/\hbar \omega _{ LO} $ is shown, where $E $ is the electron
(hole) energy, counted from energy $ \varepsilon_\ell ^{ e (h)}
$, appropriate to a size quantization energgy level $ \ell $.

The polaron states correspond to crossing points of terms. Twofold
polarons, appropriate to crossing only of two terms, are
designated by black circles. Let us consider some point of
  terms crossing, to which  an integer $j $ corresponds (see
(1)). Let $n $ is the Landau quantum number of a level, passing
through  the given point of terms at $N=0 $ (see Fig.1). Then the
conditions should be carried out
\begin{equation}
\label{107} 2j > n\geq j
\end{equation}
for a twofold polaron existence. It is easy to see that  to
 $j=1 $ one twofold polaron (designated by the
letter $A $) corresponds. Two twofold polarons $D $ and $E $
correspond to  $j=2 $ (i.e. $ \Omega/\omega _{ LO} =1/2 $), three
double polarons $F, K $ and $L $ correspond to $j=3 $ (i.e. $
\Omega/\omega _{ LO}=1/3 $), et cetera.

In Fig.1  polarons, located to the left of
 $ \Omega/\omega _{ LO} =1/3 $ are not designated. Above of twofold
polarons threefold polarons are located, appropriate to crossing
of three terms, more higher  fourfold polarons are located  and
so on. Number of polarons of each sort is equal $j $ at given $j
$ . Threefold polarons in bulk crystals are considered for the
first time in [28], in QWs - in [29-31].

Let us note that for crossing three and more terms in one point
the equidistance of Landau levels  is necessary. In order the
theory [31] of threefold polarons should be applicable, it is
necessary that the amendments to energy, caused by
non-parabolicity of bands or by excitonic effect, should be less
than splitting of terms. But in case of twofold polarons
infringement  of parabolisity  is not an obstacle, since crossing
of two terms all the same exists.

All above mentioned polarons correspond to the integer  $j $.
Besides in  Fig.1 there are other crossings of lines with quantum
number $ \ell $ (continuous lines), designated by empty circles.
They correspond to fractional $j $. As the terms crossed in these
points are characterized by $ \Delta N \geq 2 $, real direct
transitions between them with emitting of one phonon are
impossible. Let us name such polarons as weak polarons. As the
terms are crossed, their splitting is inevitable, but for
calculation  of splitting we need to take into account
transitions between crossing terms through virtual intermediate
states or to take into account the small two-phonon contributions
in the operator of electron-phonon interaction. In result
splittings $ \Delta E _{ weak} $ of terms in case of weak
polarons should be much less, than in a case of integer $j $. The
contributions of transitions through intermediate states in  $
\Delta E _{ weak} $ are of more high order than $ \alpha ^{ 1/2}
$ on the Fr\"ohlich dimensionless coupling constant  $ \alpha $.

At the account of two or more values of size quantization quantum
number  $ \ell $ the picture of crossing of terms becomes
considerably  complicated. Besides usual polarons, appropriate to
a level $ \ell ' $ (for example polaron $A ' $), occur "combined"
polarons for which the electron-phonon interaction connects two
electronic levels with different numbers $ \ell $. The Landau
quantum numbers  may coincide or be  different [32,33]. In Fig.1
for an example three terms concerning  to quantum number $ \ell '
$ (dot-and-dash lines) and a position of two combined polarons $P
$ and $Q $ are shown. In Fig.1 it would be necessary to carry out
more dot-and-dash lines and to obtain the greater number of
combined polarons. However, it
 would strongly complicate figure. For an example in Fig.1
 the polaron $R $ is
designated, which is combined and weak. Interesting feature of
combined polarons is that that the appropriate resonant values of
magnetic fields depend on distance $ \Delta\varepsilon =
\varepsilon _{ \ell '} -\varepsilon_\ell $ between size quantized
levels $ \ell $ and $ \ell ' $ and hence on QW depth and width.
Really, with the help of Fig.1 it is easily to obtain
\begin{eqnarray}
\label{108} ( \Omega/\omega _{ LO}) _P=X_P=1- (\Delta
\varepsilon) /\hbar \omega _{ LO},\nonumber\\ ( \Omega/\omega _{
LO}) _Q=X_Q=1 + (\Delta \varepsilon) /\hbar \omega _{ LO}.
\end{eqnarray}

One more kind of combined polarons [32] is not represented in
Fig.1, as it exists only under the condition
\begin{equation}
\label{109} \Delta\varepsilon = \hbar\omega _{ LO},
\end{equation}
when, for example, terms $ \ell ', n=0, N=0 $ and $ \ell, n=0, N=1
$ coincide at any magnitudes of a magnetic field. For performance
of the resonant condition (109) a certain distance between levels
$ \ell $ and $ \ell ' $ is required, what may be reached only by
selection of the QW width and depth.

In order the picture of Fig.1 should be applicable it is necessary
that the distances between the neighbour levels $ \ell, \ell -1,
\ell +1 $ should be much more, than  $ \Delta E $
 of polaron splittings. Since the distances between levels
decrease with growth of the QW width $d $, the restriction from
above on $d $ is imposed ( Numerical estimations see in [34]
(Fig.2 and 3).).

\section{Results of numerical calculations}

It follows from above mentioned classification  of magnetopolarons
that the technique of calculation of light reflection and
absorption  stated in section 7 is inapplicable only to combined
polarons in a QW, since in case of twofold combined polarons in
each of polaron states $a $ or $b $ the states with different
numbers $ \ell $ and $ \ell ' $ of size quantization "are mixed"
and as a result the expression (8) is inapplicable. To all other
twofold polarons  - usual (black circles) and weakened (empty
circles) basic formulas (84) and (85) are applicable.

As excitation 1 a pair consisting from a hole and a state $a $ of
polaron appears, as excitation 2 a pair of a hole and state $b $
appears. In narrow QWs characterized by an inequality $ \kappa
d\ll 1 $, light interacts only with those pairs for which the
size quantization numbers  coincide, i.e. $\ell_e =\ell_h $. Let
us name a creation of such EHPs by permitted transitions. For
wider QWs (when $ \kappa d\geq 1 $) forbidden transitions, for
which $ \ell_e\neq \ell_h $, appear.

Since energies $ \varepsilon_\ell^e $ and $ \varepsilon_\ell^h $,
determined in (21), depend on numbers $ \ell $, permitted
transitions with indexes $ \ell_e =\ell_h=1 $ and $ \ell_e
=\ell_h=2 $ are separated on frequency on the value
\begin{equation}
\label{110} D _{ eh} =3\hbar\pi^2/2\mu d^2.
\end{equation}
For example, for $GaAs $, using parameters [26] $m_e=0.065 \, m_0,
m_h=0.16 \, m_0 $ and QW width $d=150\AA $, we obtain
\begin{equation}
\label{111} \hbar D _{ eh} =0.11 eV.
\end{equation}

For wide QWs in an interval between the permitted transitions $
\ell_e =\ell_h=1 $ and $ \ell_e =\ell_h=2 $ there are two
forbidden transitions at $ \ell_e=1, \ell_h=2 $ on distance
$3.2\cdot 10 ^{ -2} eV $ from transition $ \ell_e =\ell_h=1 $ and
at $ \ell_e=1, \ell_h=2 $ on distance $7.9\cdot 10 ^{ -2} $ from
transition $ \ell_e =\ell_h=1 $. To each of transitions at
resonant value $H _{res} $  there corresponds a doublet of close
located maxima of light reflection and absorption. Comparative
intensity of reflection and absorption for cases permitted and
forbidden transitions is shown in Fig.3-5.
\begin{figure}
\includegraphics[]{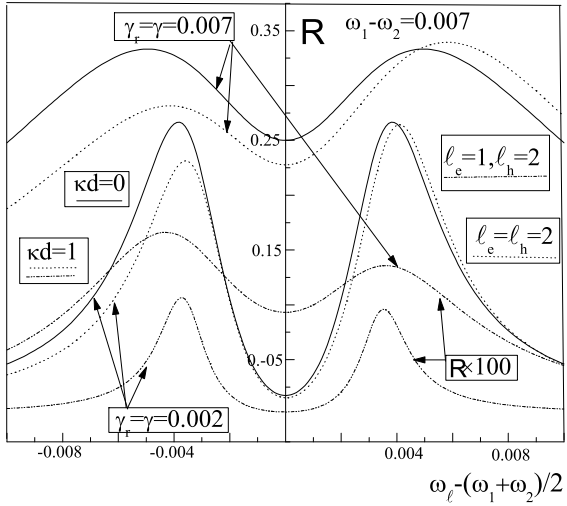}
\caption[*]{\label{Fig4a.eps}Same that in Fig.3 under condition $
\gamma_r =\gamma $.}
\end{figure}

Let us show still numerical estimations. In [34] the values $
\Delta E $ of polaron splittings are calculated for some sorts of
ordinary double polarons (see Fig.1) and some substances of QWs
 and barriers.

  Dependencies $ \Delta E $ on a QW
width $d $ are constructed.

At $d $ from $150\AA $ up to $300\AA $ $ \Delta E \approx 5\div
7\cdot10 ^{ -3} eV $ for $ \ell_e =\ell_h=1 $ and $ \ell_e
=\ell_h=2 $.  $ \hbar\gamma_r $ (according to (26) with use
parameters of $GaAs $ from [26]) is equal
\begin{equation}
\label{112} \hbar\gamma_r\simeq 5.35\cdot 10 ^{ -5} (H/H _{ res})
eV
\end{equation}
(see also [21]). At $H=H _{ res} \quad \gamma _{ ra} = \gamma _{
rb} = \gamma_r/2. $ In Fig.2 the dependence of  $ \tilde{ \gamma}
_r $ and radiative energy shifts $ \Delta $ on QW width $d $ for
two permitted transitions $ \ell_e =\ell_h=1 $ and $ \ell_e
=\ell_h=2 $ and two forbidden transitions $ \ell =1, \ell =2 $
and $ \ell =2, \ell =1 $ is represented;  results for two
forbidden transitions precisely coincide, since according to (9)
and (10) the function $ \phi_\chi (z) $ does not vary at
replacement of an index $ \ell_e $ by $ \ell_h $ and vice verse.
 \footnote{ It is right only in
approximation of indefinitely deep QWs, which is used at
construction of figures.} At $H=H _{ res} $ the non-radiative
damping $ \gamma_a =\gamma_b $, but their values are unknown, in
[21] an attempt is made only to estimate these values from below.
\begin{figure}
\includegraphics[]{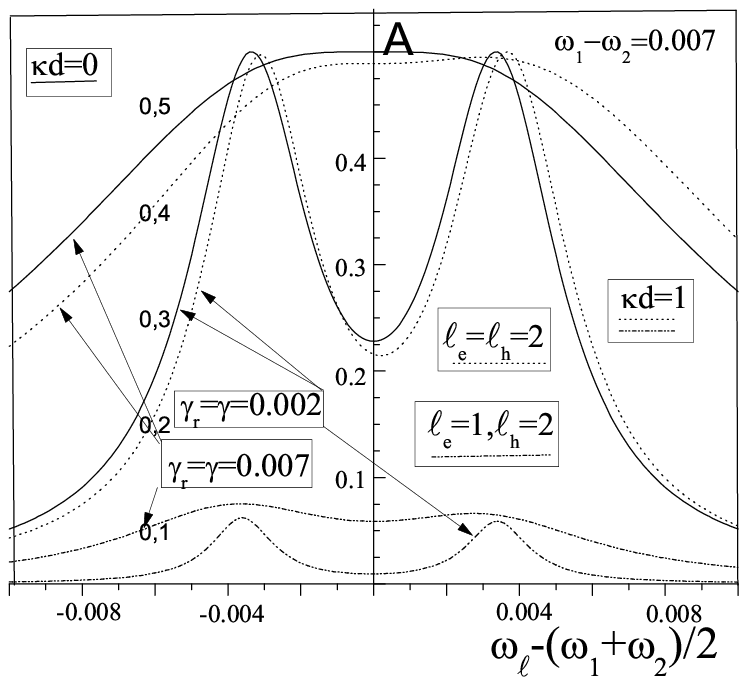}
\caption[*]{\label{Fig4b.eps}Same that in Fig.4 under condition $
\gamma_r =\gamma $.}
\end{figure}

Fig.3 corresponds  approximately to above mentioned numerical
estimations $ \Delta E $ and $ \gamma _{ r1} = \gamma _{ r2} $ at
$H=H _{ res} $. \footnote{All parameters and frequencies in
Fig.3-5 are given in arbitrary units, because the expressions (84)
and (85) contain only ratios of these values.} Two arbitrary
values $ \gamma_1 =\gamma_2 =\gamma$ are used: $\gamma =0.001 $
and $ \gamma =0.007 $, both of them exceed $ \gamma _{ r1} =
\gamma _{ r2} =2.7\cdot 10 ^{ -5} $. So  Fig.3 corresponds to a
case $ \tilde{ \gamma} _{ r1 (2)} \ll\gamma _{ 1 (2)} $. It is
visible that ${ \cal R} \ll 1, { \cal A} \ll 1 $ and ${ \cal R}
\ll{ \cal A} $  for permitted and for forbidden transitions.

Comparing the used polaron splittings $ \Delta E _{ res} $ and $
\gamma_r $ for  Fig.3 we find that  $ \gamma_r $  is two orders
less than  $ \Delta E _{ res} $. But it is right only for usual
twofold
 polarons (black circles in  Fig.1). In case of weak
polarons (the empty circles) polaron splittings are much smaller.
 Figs.5, 6 and 7, 8 may concern to weak polarons, when  $
\Delta E _{ res} $ and $ \gamma_r $ may be comparable. In Figs.5,
6  dependencies  ${ \cal R} (\omega_\ell) $ and ${ \cal A}
(\omega_\ell) $ are represented in a case
 $\gamma_{r1}=\gamma_{r2}=\gamma_1=\gamma_2$. For permitted transition at
equality of radiative and non-radiative damping ${ \cal R} $ and
${ \cal A} $ reach the greatest magnitudes comparable among
themselves and comparable with unit, what is visible in  Figs.5, 6
. As to forbidden transitions, for them the condition $ \tilde{
\gamma} _{ r1 (2)} \ll\gamma _{ 1 (2)} $ is satisfied and
appropriate  ${ \cal R} $ and ${ \cal A} $ are small.

At last, the Figs.7, 8 correspond to an inequality $ \gamma_r\gg
\gamma $, for which the most interesting results are obtained. In
both cases $ \gamma _{ r1} = \gamma _{ r2} =0.002 $ and $ 0.01 $
for a permitted transition on Figs.7, 8  we obtain results
appropriate to (99), i.e. equal to zero reflection in a point $
\omega_\ell = (\omega_1 +\omega_2) /2 $ and total reflection $
({\cal R} =1) $ in points $ \omega_\ell =\omega _{ d1} $ and $
\omega_\ell =\omega _{ d2} $, which position depends on a QW
width.

To the forbidden transition in  Figs.7, 8 at $ \kappa d=1 $
  large  ${ \cal R} $ correspond in maxima. It occurs because
   $ \tilde{ \gamma} _{ 1 (2)} \ll \gamma _{ r1 (2 _)} $ and
 $ \gamma\simeq 10 ^{ -4} $ are comparable. But peaks are very narrow,
since
 $\tilde{\gamma}_{1(2)}\ll\omega_1-\omega_2$ and
$ \gamma _{ 1 (2)} \ll\omega_1-\omega_2 $. In  Fig.8, where
dependencies  ${ \cal A} (\omega_\ell) $ are represented, to
forbidden transition at $ \kappa d=1 $ the same high and narrow
peaks correspond for the same reason.
\begin{figure}
\includegraphics[]{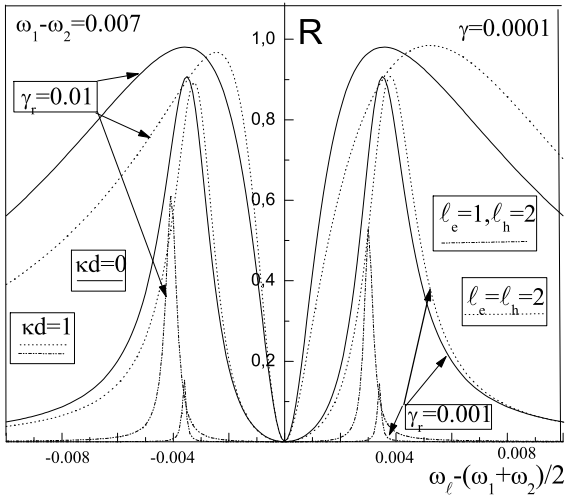}
\caption[*]{\label{Fig5a.eps}Same that in Fig.3 under condition $
\gamma_r\gg\gamma $.}
\end{figure}

\section{Conclusion}

The dimensionless  reflection ${ \cal R} $ and absorption ${ \cal
 A} $ are calculated  at normal incidence of monochromatic light on a QW surface,
which width is comparable to length of a light wave. It is
considered  a resonance of stimulating radiation with two close
located energy levels in a QW, which are two terms of a system
consisting from  a magnetopolaron and hole. Results are compared
with what were obtained earlier in [21] for narrow QWs (which
width is much less than a light waves length).

Firstly, it is shown that if a light wave length is comparable to
a QW width, height and form of peaks of reflection ${\cal R}$ and
absorption  ${\cal A}$ begin to depend on a QW width.

 Secondly, it is established, that in case of wide QWs appears
 an interaction of light with excitations, which are characterized
 by various size quantization numbers of electrons and holes ($ \ell_e\ne \ell_h $).
 Dependence of ${\cal R} $ and ${\cal A} $
 on a QW width $d $  is various for cases
 $ \ell_e =\ell_h $ and $ \ell_e\ne \ell_h $.

 Thirdly, it is shown that in case of wide QWs
 an original behaviour  of reflection ${\cal R} $  on stimulating light
  frequency $ \omega_\ell $ is preserved, if non-radiative
  broadenings of excitations
    are much less than radiative broadenings.
  There exist two points
 $ \omega_\ell =\omega _{ d1} $ and $ \omega_\ell =\omega _{ d2} $
 of total reflection
 (${\cal R} =1 $) and one point $ \omega_\ell =\Omega_0 $  between them
 of total transmission (${\cal R} =0 $). However, the
 position of points
 $ \omega _{ d1} $ and $ \omega _{ d2} $ depends on a QW width.

 All sequence of processes of absorption and reradiation
  of light quanta is taken into account, what means an exit outside
  the perturbation theory on a coupling constant of light and
 electrons.

 It is shown that the perturbation theory  is unsuitable, when
  radiative and non-radiative broadenings  are comparable.
\begin{figure}
\includegraphics[]{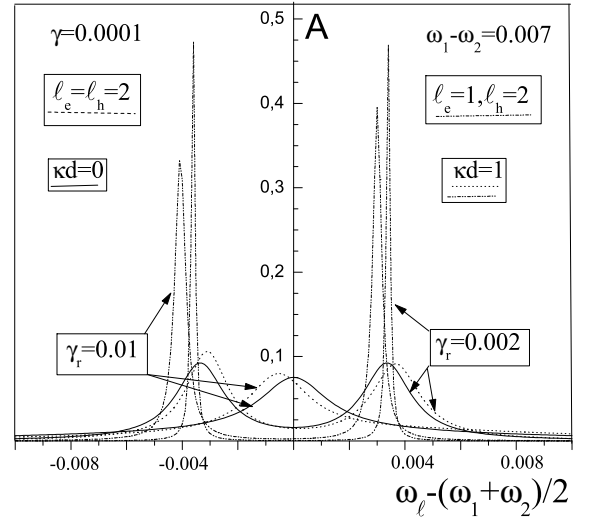}
\caption[*]{\label{Fig5b.eps}Same that in Fig.4 under condition $
\gamma_r\gg\gamma $.}
\end{figure}
\section{Acknowledgements}
This work has been partially supported by the Program "Solid State
Nanostructures Physics" and by the Russian scientific-educational
programme ``Integration''. The authors are grateful to M. Harkins
for critical reading of the manuscript.

\end{document}